# One step synthesis of $SmO_{1-x}F_xFeAs$ bulks with Tc = 54.6 K: High upper critical field and critical current density

Yanwei Ma*, Zhaoshun Gao, Lei Wang, Yanpeng Qi, Dongliang Wang, Xianping Zhang

Key Laboratory of Applied Superconductivity, Institute of Electrical Engineering, Chinese Academy of Sciences, P. O. Box 2703, Beijing 100190, China

## Abstract

A safe, simple and easily scaleable one-step sintering method is proposed to fabricate newly discovered superconductors of $SmO_{1-x}F_xFeAs$. Superconducting transition with the onset temperature of 54.6 K and high critical fields $H_{c2}(0) \geq 200$ T were confirmed in $SmO_{1-x}F_xFeAs$ with x = 0.3. At 5 K and self field, critical current density *Jc* estimated from the magnetization hysteresis using the whole sample size and the average particle size reached $8.5\times10^3$ and $1.2\times10^6$ A/cm$^2$, respectively. Moreover, the *Jc* exhibited a very weak dependence on magnetic field. Microstructural characterizations revealed that the whole sample *Jc* improvement could be achieved by either perfect texture or optimization of fabrication process in this strongly-layered superconductor. Our results clearly demonstrated that one-step synthesis technique is unique and versatile and hence can be tailored easily for other rare earth derivatives of REFeAsO superconductors.

* Author to whom correspondence should be addressed; E-mail: ywma@mail.iee.ac.cn

## Introduction

The discovery of superconductivity at 26 K in $LaO_{1-x}F_xFeAs$ compound [1] opens a new way to search for novel superconductors with higher Tc. The rather higher Tc of the oxypnictide superconductor brings new hope of both finding applications of superconducting compounds and finding new superconducting materials. The new superconductors have the general formula REFeAsO, where RE is a rare earth element [2-6], in which superconductivity is confined to the Fe-As layers and the charge carriers are provided from the doping of fluorine or alternatively by depletion of oxygen from the RE-O planes [1]. Most importantly, recent studies found that the newly discovered oxypnictide superconductor usually shows exceptionally high upper critical field ($H_{c2}$) [7-9].

However, up to now the REFeAsO superconductors are only prepared by a limited number of research groups because of the very complicated synthesis route, which requires either the high-pressure sintering process with pressure as high as 6 GPa at 1250ºC [5,6] or complex two step synthesis of pellet samples sealed in an evacuated quartz tube for annealing [1-4]. Furthermore, there is the toxicity and volatility of arsenic to consider, sometimes a sealed quartz tube of arsenic exploded during annealing. In order to simplify the fabrication process and to avoid the explosion accident, recently we have successfully synthesized La iron oxypnictide wires with Tc of ~25 K by employing a powder-in-tube route [10]. Here, we report the preparation of high-density $SmO_{1-x}F_xFeAs$ bulk samples using our newly developed one-step sintering technique. We performed XRD, SEM and magnetization measurements in applied magnetic fields in order to evaluate microstructure and electromagnetic properties.

## Experimental

Polycrystalline bulk samples with nominal composition $SmO_{1-x}F_xFeAs$ (x=0.3) were synthesized by conventional solid state reaction using high-purity Sm filings, As pieces, $SmF_3$, Fe and $Fe_2O_3$ powders as starting materials. The raw materials were thoroughly ground and encased into pure Ta tubes (One end of the tube was sealed). After packing, the other tube end was crumpled, and this tube was subsequently rotary swaged and sealed in a Fe tube. It is note that the grinding and packing processes were carried out in glove box in which high pure argon atmosphere is filled. The samples were then annealed at 1160 $^{o}$C for 40 hours. The high purity argon gas was allowed to flow into the furnace during the heat-treatment process to reduce the oxidation of the samples. The sintered samples were taken out by breaking the Ta tube.

Phase identification and crystal structure investigation were carried out using x-ray diffraction (XRD). Microstructural observations were performed using scanning electron microscope (SEM). Resistivity measurements were performed by the conventional four-point-probe method using a Quantum Design PPMS. Magnetization of the samples was measured by a SQUID magnetometer. The critical current density was calculated by using the Bean model.

**Results and discussion**

Figure 1 shows the X-ray diffraction pattern for a $SmO_{0.7}F_{0.3}FeAs$ sample annealed at 1160$^{o}$C. It is found that almost all main peaks can be indexed by a tetragonal structure with a = b = 3.927 Å, c = 8.482 Å. The two second peaks marked can be indexed to the structure of impurity phase SmAs and SmOF. Worth noting that these impurity phases could be reduced by optimizing the heating process and stoichiometry ratio of start materials.

The temperature dependence of resistivity for the $SmO_{0.7}F_{0.3}FeAs$ bulk is shown in Fig.2. The sample revealed metallic behavior where resistance decreases with decreasing temperature until it drops at 54.6 K The residual resistivity ratio RRR=(300 K)/(55 K) = 4.45. The broad transition is quite similar to what has been seen in a SmFeAsOF samples as synthesized by two-step method [11], which may be due to the phase inhomogeneity and more second phases in the sample. It should be noted that the above mentioned Tc value is quite comparable to those of bulks fabricated by the high pressure or the common two-step synthesis [11-12], indicative of effectiveness of our one-step fabrication method. The inset of Fig.2 displays the temperature dependence of magnetization in zero-field cooling (ZFC) and field cooling (FC) conditions for the $SmFeAsO_{0.7}F_{0.3}$ bulk sample at 20 Oe. It can be observed that the sample shows a superconducting transition at ~50 K. The large difference between the ZFC and FC curves of the sample suggests that the material has a fairly large flux pinning force resulting in the trapping of magnetic flux in the field cooling condition.

Magnetization loops M(H) were measured with a PPMS system at different temperatures as shown in the inset of Fig. 3. It can be seen that the magnetization loops show a strong ferromagnetic background, which is probably due to unreacted Fe or $Fe_2O_3$ impurity phase, as already observed by Flukiger et al. [8]. Large M(H) hysteresis loops imply either very strong flux pinning or a well connected superconducting state. Figure 3 shows the magnetic field dependence of the critical current density Jc derived from the hysteresis loop width using the extended Bean model $Jc = 20\Delta M/Va(1-a/3b)$ taking the full sample dimensions. In such a case, we can estimate a lower bound of $Jc = 8.5\times10^3$ A/cm$^2$ at 5 K in zero field, which is in a good agreement with the whole-sample current density result for the $SmFeAsO_{0.85}$ bulk prepared by the HP method [12]. However, if we assumed the current flows only within the grains ($Jc = 30\Delta M/R$, R is the average grain size), the Jc based on the individual grains (assuming grain size is about 10 μm estimated from SEM observations) is about $1.2\times10^6$ A/cm$^2$. It should be noted that the Jc has a very weak dependence on the applied field, very similar to those values found in Nd or Sm based oxypnictides. [8,9,13-15]

In order to get more information about the microstructures of $SmFeAsO_{0.7}F_{0.3}$ samples, SEM was performed, as shown in Fig. 4. From this figure, it is clear that the bulk sample seems denser though there are several voids observed. Clearly, the sample consists of plate-like grains of the superconducting phase with a size of ~10 μm, as shown in Fig.4 (a). The EDX results suggested that superconducting grains are compositionally homogeneous, at least within the limits of SEM-EDX analysis. Experimental data also revealed that the sample contains particle-like impurity phases, identified by EDX perhaps as $Sm_2O_3$, SmAs phases or unreacted Fe or $Fe_2O_3$. Fe or $Fe_2O_3$ inclusions are likely to be the source of the large ferromagnetic background seen on the magnetization curves. In addition, one outstanding feature as depicted by high magnification SEM images is that the superconducting plate-like grains exhibit a delaminated layer structure, as shown in Fig.4 (b). Multiple layers forming a large grain can be easily observed in all the samples, indicative of a layer growth mechanism for the formation of [(Sm-O-F) (FeAs)] phase. This granular behavior is in analogy to high Tc Bi-based compounds.

We carried out the resistance versus temperature measurements on the bulk sample in various magnetic fields by the four-probe resistive method. The onset temperature decreases very slowly with increasing the magnetic field, however, the Tc(0) drops quickly to low temperatures. The 10% and 90% points on the resistive transition curves were used to define the $H_{irr}$ and $H_{c2}$. The upper critical fields are determined in this way and shown in Fig. 5. It is clear that the irreversibility field is rather high comparing to that in $MgB_2$. The curves show a positive curvature very near to Tc similar to the $MgB_2$ case. The slope of $H_{c2}$ near Tc (99% Rn from R-T), $dH_{c2}/dT|_{Tc}$, is ~5.3 T/K. The corresponding $H_{c2}(T=0)$ value derived from the Werthamer-Helfand-Hohenberg formula is-0.693Tc $(dH_{c2}/dT)_{Tc}$= 200 T. These high values for upper critical fields indicate a very perspective application of this new superconductor.

SmFeAs(O,F) superconductors with critical temperatures Tc = 54.6 K and very high critical fields ≥200 T have been prepared by one-step sintering technique, demonstrating that this synthesis method developed by us is safe, easy and very promising to fabricate high quality REFeAs(O,F) high temperature superconductors. However, the global Jc values of $8.5\times10^3$ A/cm$^2$ at 5 K obtained is significantly lower than that seen in random bulks of $MgB_2$ which generally attained $10^6$ A/cm$^2$ at 4.2 K [16]. The main reason may be ascribed to the second impurity phases or the presence of weak links between the grains. As supported by XRD and SEM/EDX results, major impurity phases such as $Sm_2O_3$, SmAs phases and unreacted Fe or $Fe_2O_3$ were observed in our present samples. These macroscale impurity phases can significantly reduce percolating current path and limit the globe Jc. Even in such case, very high intragrain $1.2\times10^6$ A/cm$^2$ and wide magnetization hysteresis loops indicated that strong intragrain vortex pinning was existed in the sample [13]. Based on the present results, it is believed that great improvement in the globe Jc is expected upon either optimization of processing parameters or achieving high grain alignment in analogy to high $T_c$ Bi-based cuprates.

**Conclusions**

In summary, we have prepared $SmO_{1-x}F_xFeAs$ bulk materials using one-step sintering method. High temperature superconductivity with the onset temperature of 54.6 K and higher critical fields ≥200 T were confirmed. The Jc estimated from the magnetization hysteresis using the sample size was $8.5\times10^3$ A/cm$^2$ at 5 K and self field. The Jc has much potential to be enhanced by perfect texture or optimization of fabrication process in this new superconductor. Our results clearly demonstrated that one-step synthesis technique is unique and versatile and hence can be tailored easily for other rare earth derivatives of REFeAsO superconductors.

**Acknowledgement**

The authors thank Chenggang Zhuang, Xiaohang Li, Haihu Wen, Liye Xiao and Liangzhen Lin for their help and useful discussion. This work is partially supported by the Beijing Municipal Science and Technology Commission under Grant No. Z07000300700703, National ‘973’ Program (Grant No. 2006CB601004) and National ‘863’ Project (Grant No. 2006AA03Z203).

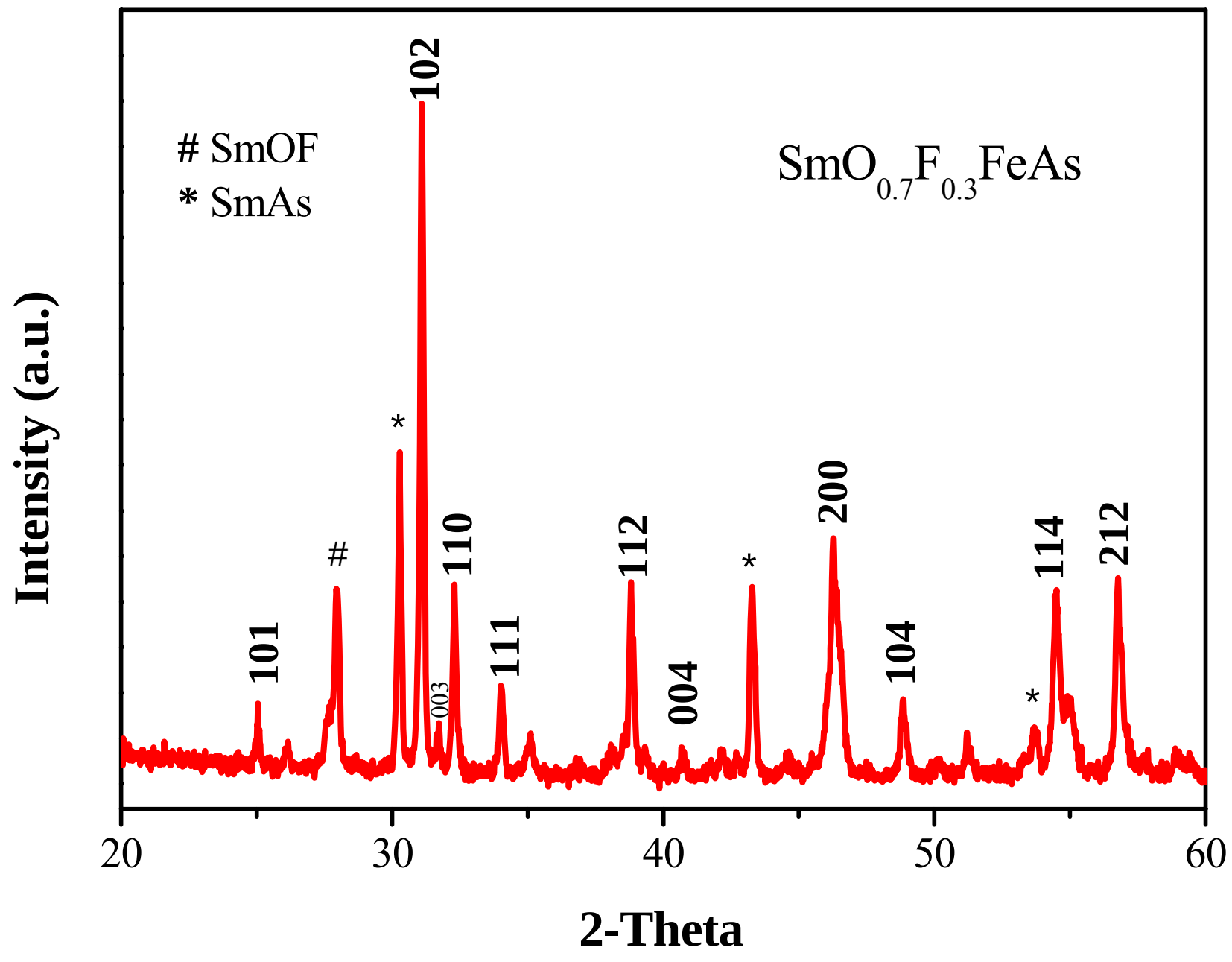

**Figure 1** X-ray diffraction pattern for a sample with nominal composition $SmO_{1-x}F_xFeAs$ (x=0.3).

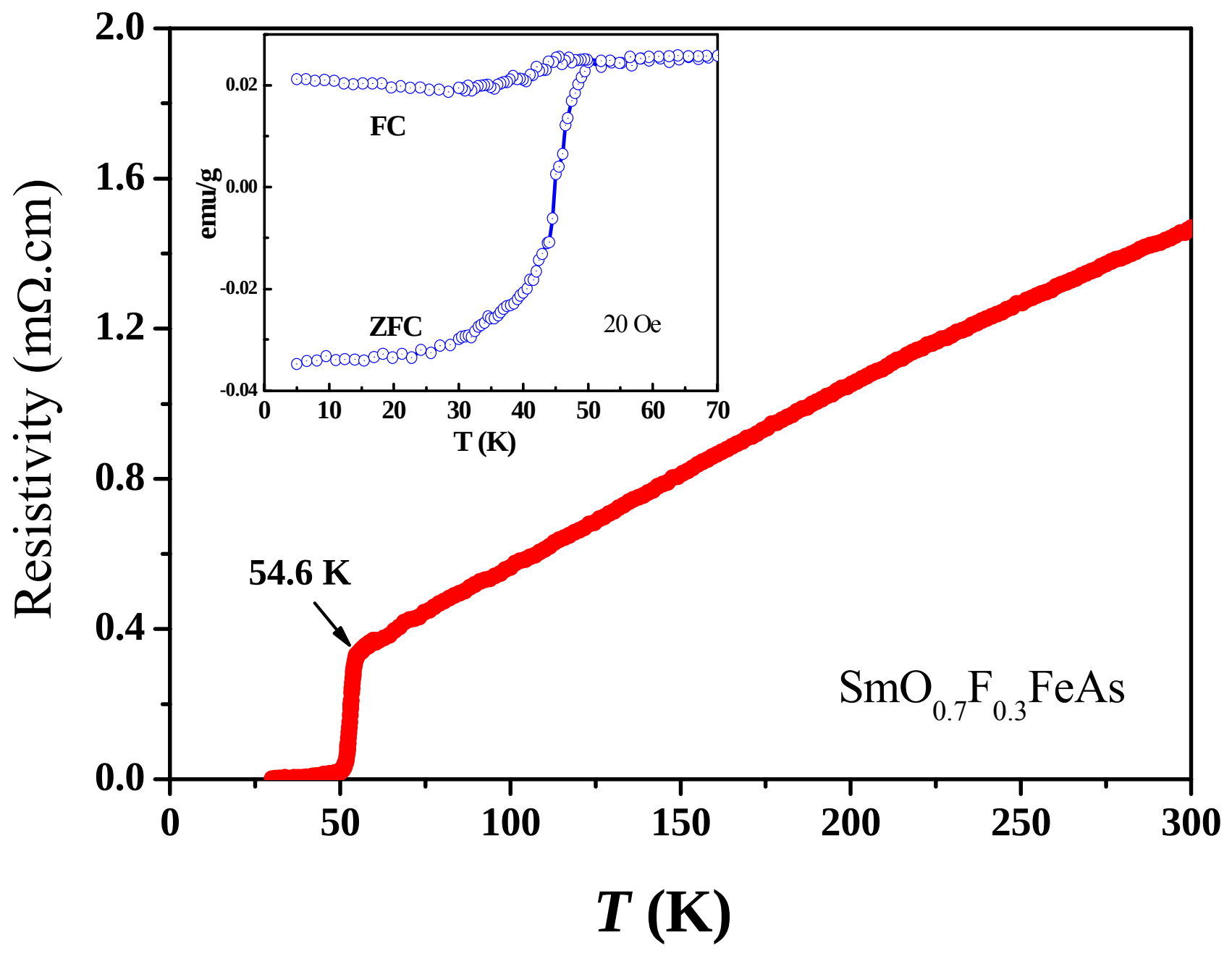

**Figure 2** Temperature dependence of resistivity for the $SmO_{0.7}F_{0.3}FeAs$ sample. Inset: the temperature dependence of magnetic susceptibility measured at 20 Oe.

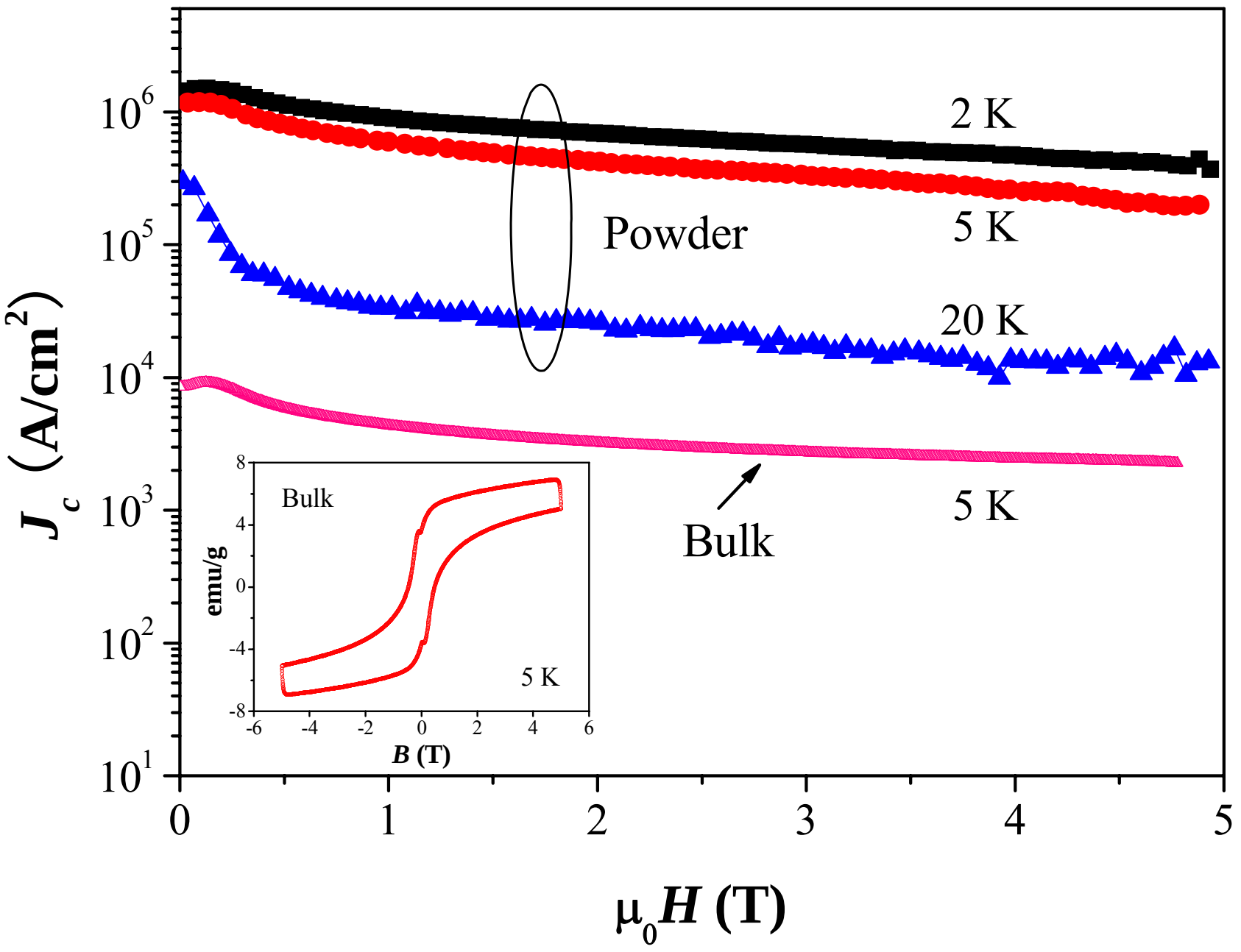

**Figure 3** Magnetic field dependence of the critical current density Jc at different temperatures for the powder and the bulk samples. The inset shows the bulk *M-H* loop at 5 K with the ferromagnetic background.

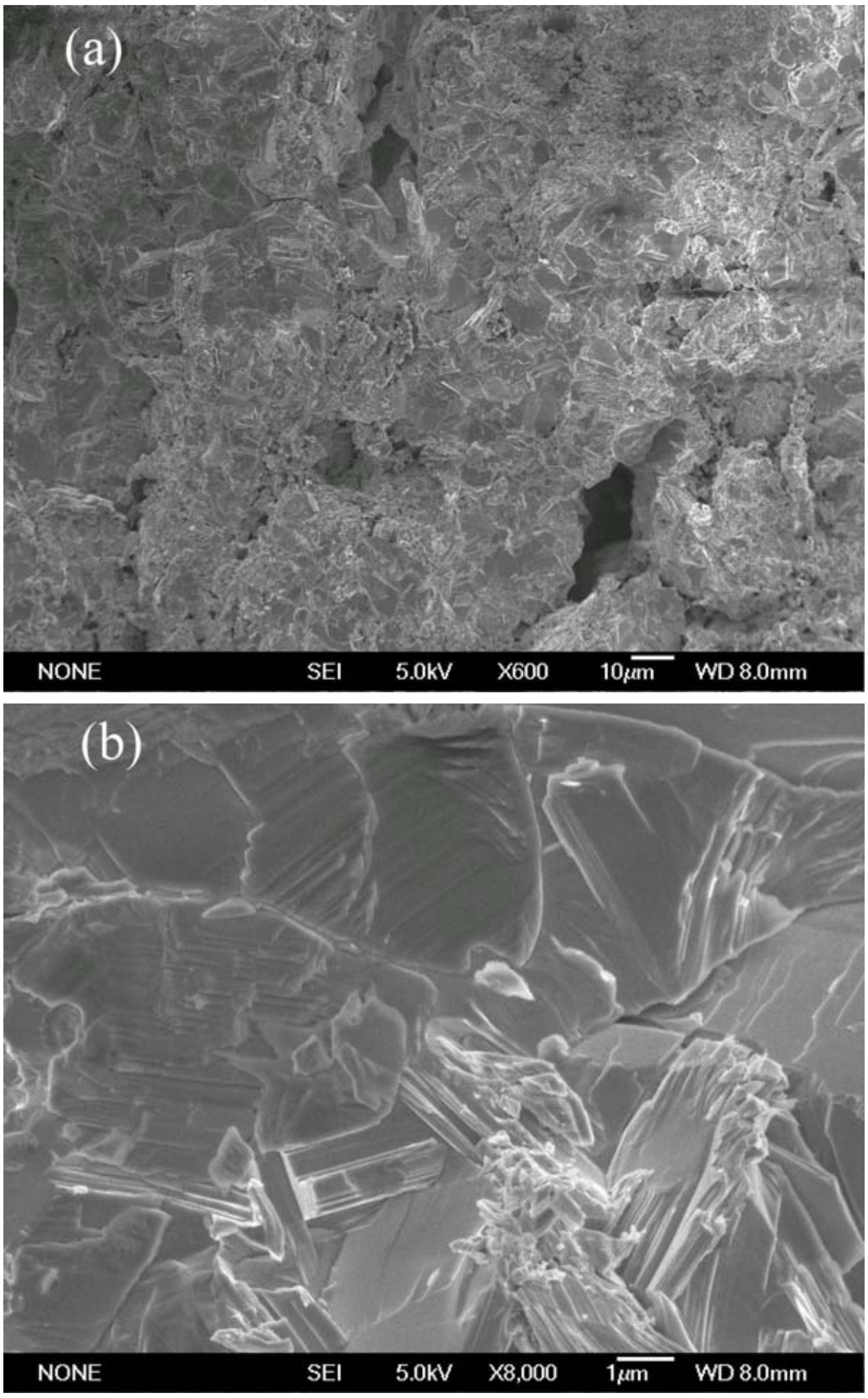

**Figure 4** Typical SEM images of the fractured surfaces of the $SmFeAsO_{0.7}F_{0.3}$ samples. (a) Low magnification, (b) Enlarged view of plate-like grains.

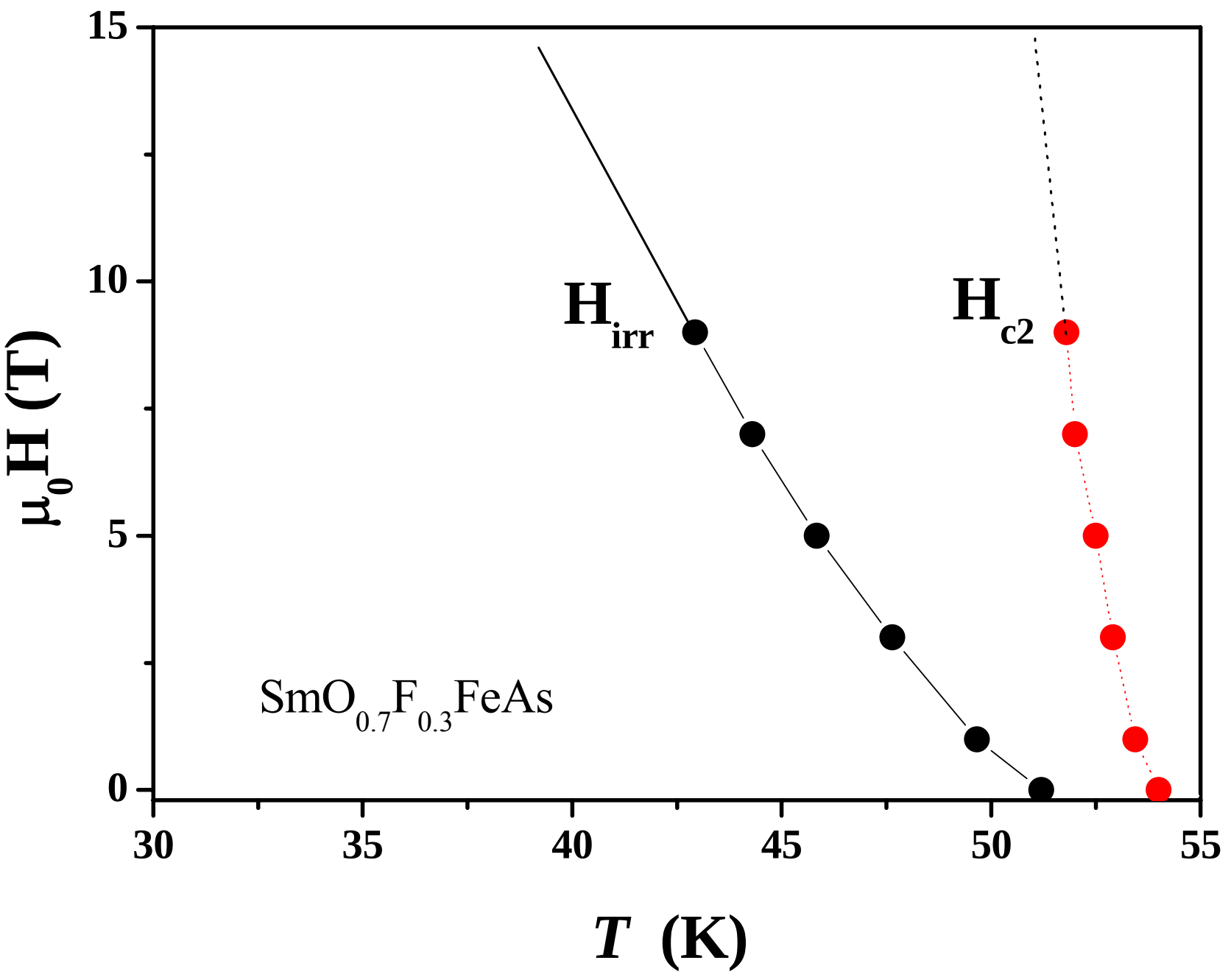

**Figure 5** The upper critical field line $H_{c2}$ and $H_{irr}$ as a function of the temperature for the $SmFeAsO_{0.7}F_{0.3}$ samples. The $H_{irr}$ and $H_{c2}$ values were defined as the 10% and 90% points of the resistive transition, respectively.